\documentclass[10pt, twocolumn]{article}
\usepackage{graphicx}
\usepackage{url}

\title{Security and Privacy Issues in Cloud Storage}
\author{Norah Asiri\\norah.m.asiri@outlook.com}

\begin{document}
\maketitle
 	\paragraph{Abstract—}
Even with the vast potential that cloud computing has, so far, it has not been adopted by the consumers with the enthusiasm and pace that it be worthy; this is a very reason statement why consumers still hesitated of using cloud computing for their sensitive data and the threats that prevent the consumers from shifting to use cloud computing in general and cloud storage in particular. The cloud computing inherits the traditional potential security and privacy threats besides its own issues due to its unique structures. Some threats related to cloud computing are: the insider malicious attacks from the employees that even sometime the provider unconscious about, the lack of transparency of agreement between consumer and provider, data loss, traffic hijacking, shared technology and insecure application interface. Such threats need remedies to use its features in secure way. In this review, we spot the light on the most security and privacy issues which can be attributed as gaps that sometimes the consumers or even the enterprises are not aware of. We also define the parties that involve in scenario of cloud computing that also may attack the entire cloud systems. We also show the consequences of these threats. 

\section{Introduction}
\begin{figure}
	\centering
	\includegraphics[width=0.7\linewidth]{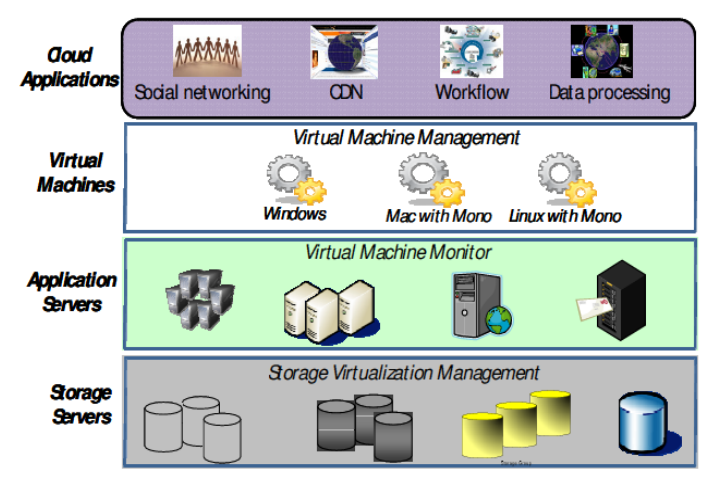}
	\caption{Typical data center}
	\label{fig:screenshot001}
\end{figure}

Cloud Computing is a new technology of delivering computing resources which conveys a huge influence to IT society. It provisions on-demand services based on Internet infrastructure, where the resources are configurable. With cloud computing systems, people run their businesses on shared data center as they just plug in as utility using typical data center as shown in Figure \ref{fig:screenshot001}. Cloud computing is not just a consumer's applications, but also for business applications where an enterprise runs all kinds of applications including custom built applications. Generally, the cloud computing systems are more scalable, secure and reliable than the vast majority of other systems. Obviously, doubling data of organization makes it feels pressure. The demand of extra low cost storage is excessively required. Cloud storage may solve this problem by providing these organizations with extra scalable storage but with potential storage associated risks \cite{subashini2011survey}.

\paragraph{Cloud storage:} It is a model of storing data on hard drive or shared pool that is hosted by a third party through the Internet. It permits consumers to access their data virtually anywhere.
\paragraph{Storage-as-a-Service (StaaS):} It allows consumers to consign and store their data remotely, and enables them to access data any time from any place. StaaS ties storage with specific data format and applications. Wide range of cloud storage offers general-purpose storage like Dropbox. Yet, it has not been widely utilized. It should be considered the cloud storage is not widely covered and explained in services models of cloud computing \cite{wang2011privacy}.

\section{Cloud deployment models}
\begin{itemize}
	\item Public Cloud: A public cloud is a cloud that offers a shared infrastructure from personal storage to complex suite of enterprise services among several untrusted consumers. It usually consists of data center owned by service provider who manages the infrastructure and storage systems and sell these services. From the consumer's view, it is an advantage because the infrastructure shared across many users and everyone pays less than when they run their own cloud.
	
	\item Private Cloud: It is a data center that is dynamically provisioned and the infrastructure is dedicated to specific organization according to its requirements. In private cloud, the organization's data stays in its data center and will be always available. The integration with this level is much easier. The consumer should pay for all these privileges.
	
	\item Community Cloud: In the community cloud, the provider who hosts the services could be settled within the organization or could be a third party. That means, the cloud can be all on or off premises. Generally, it could serve a particular group or whole organization.

	\item Hybrid Cloud: Hybrid cloud is the combination of two or more of cloud types depending on the needs of an organization. It is all about merging the advantages of	two or more approaches of cloud kinds.
	
\end{itemize}

\section{Cloud participants}
Cloud model usually consists of four participants as follows:
\begin{itemize}
\item Cloud Provider: A cloud provider is an entity that administers the services and in charge of making them available.
\item  Cloud Consumer: A cloud consumer can be cloud service owner or cloud service consumer. Cloud service owner is the responsible entity to subscribe and if there is a change of service the cloud owner should pay the bills. Cloud service consumer is an entity (could be a person or an application) who accesses the cloud services.
\item Cloud Broker: A cloud broker is the entity who decides the suitable service to the cloud consumer.
\item Cloud Auditor: A cloud auditor is an independent entity who assesses the security and privacy of cloud service stack. 
\end{itemize}

\section{Cloud storage associated risks}
 
Using cloud storage is designed to provide and discover the capabilities of its scalability and ease of management. Once the user shifts his data to off-premises untrusted third party, this may expose his data to threat. It strongly becomes important to define any threats to confidentiality that exist in this setting since more data transferred to third party in cloud computing storage model. Generally, manage and implement such container faced some advantages and challenges. These 
advantages include:
\subsection{Advantages of cloud storage over traditional storage}
\begin{itemize}
\item Elasticity and simplicity: The consumer able to adjust the capacity of storage, and the simplicity of managing the contents of storage and getting from any 
location connected to the Internet. Also the consumer able to allow others to access data and even modify it.
\item Scalability: Cloud storage should be more convenient and offers more flexibility and enable consumer to expands and shrinks according to needed capacity. 
\item Security: Public cloud usually tends to be less secure then private cloud due to its nature and design purposes. One solution of public cloud storage security is to store data on a partition of a shared storage system. Another solution is using cryptography and accessing control over the data especially for enterprises.
\item Performance: The performance of storage infrastructure should be highly considered since the performing of data storing and recovery is mandatory elements of cloud storage.
\item Reliability: In particular, replicating data across consumer's accounts and the consumer make sure that their data is recoverable; a highly reliable storage should be constructed.
\item Ease of Management: The consumer assumes always enough space on cloud storage and always able to catch required files. Such requirements need to be managed to encounter any storage disaster and satisfies the enterprise's expectation from cloud storage deployment.
\item Ease of Data Access: Accessing data in cloud storage should be tied to access control roles but at the same time it should be easy for the verified user to access data.
\item Mobility: Employees can access information wherever they are, rather than having to remain at their desks.
\end{itemize}

\subsection{The clients and cloud storage adoption challenges}
\subsubsection{Trust in cloud storage}
Khaled M. Khan and Qutaibah Malluhi in Establishing Trust in Cloud Computing \cite{khan2010establishing} define the trust in cloud storage. Trust means "an act of faith; confidence and reliance in something that’s expected to behave or deliver as promised". The term of trusting itself is the major concept in the interactions among people. The adaption of secured and trusted cloud storage-based system is highly demanded systems due to the attacks that threaten both systems distrustful providers and users. By considering a system provider who wants to assure the user is not malicious hacker and the user to be assured the privacy and consistency, the goal of develop cloud storage succeeded. Since the interaction is a significant role in cloud systems, the mutual trust between cloud service provider and the user is consider to be main component in this physical, logical and personal controls. Demanding such kind of trust needs to a proactive approach that puts the user along with provider in controlling processes. Assuming “secure cloud” or saying “trust me” to the consumer don’t help much to boost the trust level of consumers. Since the data stored in remote third party's storage, the consumer trusts the remote system-cloud computing in this case- less than on-site stored data. Trust in cloud storage is not merely guaranteeing compensation it is more related to preventing violation against corruption of data or compromising of integrity, confidentiality or availability. Moreover, lack of transparency can affect the trust negatively. Generally, consumer always percepts that the in-site stored data is more secured than off site data. To regain the consumer's trust, the transparency and the control over the consumer's data in cloud storage should be enhanced.

\subsubsection{Immaturity of cloud storage}
Tharam Dillon and Chen Wu and Elizabeth Chang in Cloud Computing: Issues and Challenges \cite{dillon2010cloud} claim that cloud computing still in its early stages. The immaturity of cloud storage leads to unclear structure and capabilities of this storage. The uncertainty with new dramatic changes in storage technology leads to very little known about construction and architecture of such storage. The first study and analysis on the leading solution of personal cloud storage Dropbox titled with Inside dropbox: understanding personal cloud storage services just has occurred in November 2012 which means very recently \cite{drago2012inside}. Very limited studies have covered and discussed cloud storage, its structure and the risks associated with it. Ordinary consumer even if he trusts the provider or service owner, he might have no idea about the mechanism of cloud storage itself. The immaturity of cloud storage makes it difficult to meet all standards. On the other hand, the providers rush into adopting cloud computing storage system before appropriate solutions to associated issues. Hence, the gap between the adoption cloud storage and consumer is still wide. The confusion in cloud storage actually related to excited investors who rush to enter the market creates lack of standards. Such problem leads to not fully tested environment to integrate and relocate the traditional storage system with cloud storage system. The immaturity of cloud storage also leads to misunderstanding responsibilities. For example, in traditional storage system, the database management systems have been defined in details and encompassed all technologies and operations to fully improve this storage performance. Whereas for cloud storage just little known about this system and little systems have adopted by using cloud storage unlike the traditional storage system where so many system adopted based on this kind of storage \cite{dillon2010cloud}.

\subsubsection{Providers, cloud services owner and anonymous access}
Gansen Zhao et el., in Trusted Data Sharing over Untrusted Cloud Storage Providers \cite{zhao2010trusted} discuss one of the main issues that prevents adopt cloud storage for sensitive environment such as healthcare system and financial organization, is its security and privacy concerns. Answering security related questions in cloud storage is really crucial. One of the most known security challenge is provider and anonymous access and how to keep data stored in cloud storage confidential and not compromising it by any means. Threats to the consumers of cloud storage are malicious actions by the remote service provider, who is definitely is able to harm customer confidentiality and integrity. A malicious party can run and control many entities in the cloud, just by contracting for them. For example, in case of using public cloud, data owner should consign their data on untrusted cloud domain that may harm these sensitive data. Unfortunately, cloud service owner of cloud services and storage can play the role of provider and will be able to disclose the user's sensitive data. Moreover, cloud service provider has almost fully access control over the user's data which gives the providers the ability to disclose or modify data or even systems. To overcome the issues of untrusted storage and providers, users should be provided with mechanism protect them and enable them to store their data over untrusted cloud storage and not reveal it without permission to any other participants \cite{dillon2010cloud}.

\section{Security and privacy issues in cloud structure}
The biggest cloud concern in consigning data to cloud storage is data security and privacy. Consumers view about cloud computing platform is that it is less secure than the traditional network infrastructure. Though these two terms –security and privacy- are usually interleaved, the security policy ensures the privacy. Data security is commonly means the protection of data confidentiality, availability, and integrity. Whereas the data privacy means protecting the individuals data against any disclosure and misused from attackers and to feel free from all interferences. The relationship between security and privacy should be guaranteed in cloud virtual and dynamic environment especially sensitive stored data related to organizations or individuals to fully enjoy the benefits of cloud storage. 

\subsection{Data traffic and man-in the—middle, back-door and replay attack}
All existing attacks still work in cloud storage systems. Verizon Business in their Verizon Business DataBreach Investigation Report \cite{subashini2011survey}, reported of the breaches involve hacking with the following breakdown: application/service layer39\%, OS/platform layer 23\%, exploit known vulnerability18\%, exploit unknown vulnerability 5\% \cite{wu2010network}. Man-in-the-middle attack is an old attack that used against protocols and harms the privacy and authenticity of data transferring over cloud storage. It is an active attack where the attacker able to establish an independent connections and substitutes his public key of the victim. Later, if any one sends an encrypted message to the victim, the attacker would be able to read, modify and forward the message to victim. Cloud storage system should be supplied by secure techniques such as secure sockets layer (SSL), otherwise; if such techniques are incorrectly configured then the authentication may not act as estimated therefore man in the middle attacks occurred. Back-door attack used to get connected to the network and access it through allowing passing any authentication measures that would allow the attacker to bypass any accessing control roles and access stored data. For instance, images that taken from untrustworthy sources can establish a back-door bug into the cloud storage. The use of backdoor attack in cloud computing storage is 5\%. Replay attacks is the interception of victim's transmission and maliciously attacks his messages and try to send the later by impersonate the receiver. Unfortunately, the replay attack could occur even with cloud storage system while the user sends or receives data from or into cloud storage.
\subsection{Leakage in third-party compute clouds- cross-VM information leakage}
By 1970, IBM started the development of virtualization technology and it began with development of the system/360. Virtual machines(VM) is a technology of coupling the hardware and operating systems and share them among different instances that positioned on physical servers to improve the performance of these servers. These virtual instances are isolated from each other. Virtualization in cloud computing has different of types like virtualization of servers, virtualization of storage and virtualization of network. Neil MacDonald stated in Security considerations and best practices for securing virtual machines published in March 2007 “Through 2009, 60\% of production VMs will be less secure than their physical counterparts" \cite{wu2010network}. Generally, adoption virtualized environment leads to many complex security risks that have dramatic consequences \cite{rehman2014virtual}. In normal attack, with little dollars, it is able to plant malicious VM on the server as target customer is 40\% succeeded \cite{hay2011storm}. Due to the sharing of physical resources (e.g., CPU’s data caches) and multitenancy, there exist dangers when consign sensitive data to third-party cloud computing. 
Even though the provider and its infrastructure considered as trusted party, the violence and vulnerabilities of cloud could be inside the cloud management systems (e.g., virtual machine monitors). Undoubtedly, virtualized environment is a key feature of cloud computing. However, virtual machine (VM) is considered non-obvious threat and to be one of the critical concerns about security in cloud computing due to physical resources can be transparently shared between virtual machines (VMs). In \cite{fanning2011virtualization,pearce2013virtualization}, authors analyze the security violence in virtualization environment and pointed out these issues as following:

\begin{itemize}
	\item The isolation in VM is easily breakable. For example, the VM Monitor keeps track of the availability of VMs and their resource privileges. Hence, VM can 	monitor another one or even access to the host machine.
	
	\item VMs are able to be managed remotely which create critical security violence such as cross-site scripting, SQL injection.

	\item In VMs, the possibility of denial of service (DOS) attack is high. Since the resources such as memory, storage, and central processing unit are all shared in VMs with hosts, VMs are able to deny all requests to these available resources from the hosts.
	
	\item Virtual machine based Rootkit issues. Rootkit tool is a set of programs that allows the administrator-level to access the hosts. In case of Rootkit is compromising the hypervisor which allows each physical server to run several “virtual servers”, the Rootkit will able to control the entire physical machine.
	
	\item Snapshot problem. Snapshot generally allows the administrator to take a snapshot of machine in a certain point and to revert the snapshot if needed. Using such mechanism brings security issues with it retrieving passwords and abusing security policies.
\end{itemize}

\subsection{Outsourcing data and applications}
In \cite{jajodia2011privacy}, authors define the disclosing of outsource data and propose a solution. To ensure the security of confidentiality, cloud storage should be accessed only by the authorized entities. As known, cloud storage usually managed remotely and by third party to take the decision of accessing control. The contract between provider and consumer may help to prevent any unauthorized access but consumer needs to prevent the provider also from abusing data. 

\subsection{Sharing responsibility}
The main issue in cloud storage is that data resides in shared environment and responsibility and the security and privacy is shared between consumer and provider. However, other parties rather than the data owner who the only one should has full access to data are allowed to access the data even without knowledge of owner due to heterogeneous storage system that leads to compulsory shared responsibilities \cite{pearce2013virtualization}. This may lead to remove the burden of controlling data in remote storage from consumer but increase the risk to disclose his private sensitive data. Besides its low cost and huge capacity provided to entities services, controlling of accessing should be managed by only consumers and eliminate any other interventions. In current cloud storage system such crisis is inescapable but the demand of self-accessing control is increasing to preserve the security and the privacy of consumer's data.

\section{Conclusion}
Cloud computing is undeniable the future coming technology. It has tremendous capabilities of providing scalable services besides the scalable storage. Some people may think that some threats are inescapable and unavoidable since the providers own the service and have more privileges than other consumers. Moreover, though of predefined policies and signed agreements between consumers and providers the consumers' expectations not always satisfied, some barriers may happen that prevent the providers to keep their promises. One simple solution, which Milne (2010) states to use this model for UK businesses is to simple use in-house ‘‘private clouds’’ (Milne, 2010) \cite{milne2010private}. Fortunately, the solutions are available and so many techniques could handle the listed issues. We recommend using private cloud storage services for sensitive systems. If not able for the organizations or the individuals to use such deployment model, we recommend to not store the sensitive data on the cloud storage or at least encrypt them before send them remotely to cloud storage.

	\bibliographystyle{plain}

\bibliography{ref_cloud}

\end{document}